\documentstyle[12pt,amssymb,preprint,prd,aps,eqsecnum,epsf]{revtex}
\tighten

\begin{document}

\draft
\preprint{DAMTP-1998-106}
\title{Extending the Black Hole Uniqueness Theorems \\ II. Superstring Black Holes}
\author{Clive G. Wells\thanks{Electronic address: C.G.Wells@damtp.cam.ac.uk}}
\address{Department of Applied Mathematics and Theoretical Physics,
University of Cambridge, \\ Silver St., Cambridge CB3 9EW, United
Kingdom} 
\date{\today}
\maketitle

\begin{abstract}

We make use of an internal symmetry of a truncation of the bosonic sector of the superstring and $N=4$ supergravity theories to write down an analogue of Robinson's identity for the black holes of this theory. This allows us to prove the uniqueness of a restricted class of black hole solutions. In particular, we can apply the methods of the preceding paper to prove the uniqueness of a class of accelerating black holes (the Stringy Ernst solution and Stringy $C$-metric) which incorporate the possibility of the black hole accelerating within an electromagnetic flux tube. These solutions and their associated uniqueness may be useful in future instanton calculations.

\end{abstract}


\narrowtext

\def\nonum{\nonumber\samepage}
\relpenalty=10000\binoppenalty=10000

\def\be{\begin{equation}}
\def\ee{\end{equation}}
\def\beq{\begin{eqnarray}}
\def\eq{\end{eqnarray}}

\def\bd{\b\delta}
\def\sh{\mbox{$\frac12$}}
\def\D{\nabla}
\def\frc#1#2{\displaystyle\frac{#1}{#2}}
\def\sn{\mathop{\rm sn}\nolimits}
\def\cn{\mathop{\rm cn}\nolimits}
\def\dn{\mathop{\rm dn}\nolimits}
\def\sc{\sn\chi}
\def\cc{\cn\chi}
\def\dc{\dn\chi}
\def\dd{\mbox{\boldmath $d$}}
\def\e#1{\mbox{\boldmath $e$}^#1}
\def\hd{\mbox{\boldmath$*$}}
\def\b#1{\mbox{\boldmath$#1$}}
\def\k{\b k}
\def\lk{{\cal L}_K}
\def\lm{{\cal L}_m}
\def\({\left(}
\def\){\right)}

\def\ddd#1#2{\dd#1\otimes\dd#2}
\def\pp#1#2{\frc{\partial#1}{\partial#2}}
\def\ppp#1{\frac\partial{\partial#1}}
\def\df#1{\hat#1}
\def\X#1{{X^{(#1)}}}\def\Y#1{{Y^{(#1)}}}
  
\section {Superstring Black Hole Uniqueness Theorems}
\label{chap:super}

\subsection{Introduction}

In this paper we extend the black hole uniqueness theorems to the Superstring 
and $N=4$ Supergravity theories. We will be making use of the notations and conventions in~\cite{Wells}, where we proved the uniqueness of a class of accelerating black holes in Einstein-Maxwell theory. In order to make progress in the string theory we will be studying we will need 
to impose staticity rather 
than merely stationarity of the solutions, and naturally require the 
invariance of the dilaton under the action of the isometries generated
 by the Killing vectors. In addition we will only consider the case where the axionic 
field has been set equal to zero. This is consistent if we assume the 
electric and magnetic components are actually derived from two separate 
$U(1)$ gauge fields. The essential point to notice in our proof is that
 the effective Lagrangian in such a theory can be written as  the sum of two copies of that which we find for pure gravity. We will 
need to verify 
that the Weyl coordinate system may be introduced and then
 make use of Robinson's identity to establish the uniqueness result.

Firstly we will establish the uniqueness of a class of black holes 
obtained by performing an internal symmetry (the Double Ehlers' transform) to a spherically symmetric solution found by 
Gibbons \cite{NuclB}. These solutions are asymptotically Melvin's 
Stringy Universe, it thus generalizes the result of Hiscock \cite{Hiscock} for 
the Einstein-Maxwell theory. We could equally apply the theory to 
asymptotically flat 
solutions but one might feel that the uniqueness of such solutions should be 
proved under less stringent hypotheses, in particular 
Masood-ul-Alam has already
proved the uniqueness of an asymptotically flat black hole solution
in these theories \cite{Masood}.

Secondly, we turn to the Ernst solution and the $C$-metric, or rather their
 stringy variants and proceed to prove a theorem establishing their 
 uniqueness (see for the proof in Einstein-Maxwell theory~\cite{Wells}).
The solutions found here represent a generalization of those discussed by
 Dowker {\em et al.\/} \cite{Dowker}, and reduce to them when the Double
 Ehlers' transform has equal parameters. It might be noted that they do 
not agree with those previously proposed by Ross \cite{Simon}.

In Sect.~\ref{sect:cos} we introduce the spherically symmetric solution in
string theory that is the analogue of the Reissner-Nordstr\o{}m black hole.
We then perform a double Ehlers' transform to generate a new solution that
will be the object of our uniqueness theorem. In the following section, 
Sect.~\ref{sect:thyp}, we carefully state the hypotheses we need to prove
the theorem and justify the introduction of Weyl coordinates by proving that
the norm of the Killing bivector is a harmonic function on the relevant
orbit space.

In Sect.~\ref{sect:divid} we explain how Robinson's identity for the pure
gravity can be exploited to give us a tool for establishing a uniqueness
theorem in string theory and $N=4$ supergravity subject to the hypotheses
laid out in Sect.~\ref{sect:thyp}. We then complete the proof of our theorem
by presenting sufficient boundary conditions to make the appropriate
boundary integral vanish. These conditions are laid out in 
Sect.~\ref{sect:n4bcs}.

Having demonstrated how we may establish a uniqueness theorem in these theories
we go on to apply our methods to the Stringy $C$-metric and Stringy Ernst
solution. The Stringy $C$-metric is that found by Dowker {\em et al.\/} 
\cite{Dowker}. We apply the double Ehlers' transformation to derive
the Stringy Ernst solution. As in \cite{Wells} we transform coordinates to ones which have a strong
relationship to the elliptic functions and integrals. This is set out in Sect.~\ref{sect:stringyunique}. Then in 
Sect.~\ref{sect:stringybcs} we write down the relevant boundary conditions to
complete the uniqueness theorem for these solutions. Finally in the conclusion,
Sect.~\ref{sect:stringyconc}, we make a few comments on the difficulties in
generalizing the result.

\subsection{The $N=4$ Supergravity and Superstring Theories}
\label{sect:n4}

Let us turn to a truncated theory arising from the bosonic sector of the
$N=4$ Supergravity and Superstring Theories. These theories possesses a dilaton with coupling parameter equal
to unity, as well as electric and
magnetic potentials.
For simplicity we will restrict attention to the static truncation of the
harmonic map. The $N=4$ theory possesses an axionic field, and six $U(1)$ 
gauge fields that combined have an $SO(6)$ invariance. Together with a 
suitable duality rotation it is possible to reduce the theory to one 
with just two $U(1)$ gauge fields, one purely electric, the other 
purely magnetic. At this point the axion decouples and can be 
consistently set equal to zero. What remains can be written in 
terms of an effective single electromagnetic field (with both electric 
and magnetic parts), see Gibbons \cite{NuclB} for further details. 
The Lagrangian density can then be written:
\be
{\cal L}=\sqrt{|g|}\(R-2|\D\phi|^2-e^{-2\phi}F_{ab}F^{ab}\)\ .
\ee
After a dimensional reduction on a spacelike
axial Killing vector field
$m=\partial/\partial\varphi$ the density takes the form:
\be
{\cal L}=\sqrt{\gamma}\({}{}^3\!R-2\(
\frc{|\D X|^2}{4X^2}+|\D\phi|^2+\frc{e^{-2\phi}|\D\psi_e|^2}X+\frc{e^{2\phi}
|\D\psi_m|^2}X
\)\)
\label{eq:n4lag}
\ee
where
\beq
\b g&=&X\ddd\varphi\varphi +X^{-1}\gamma_{ij}\dd x^i
\otimes\dd x^j,\label{eq:hd2}\\
\dd\psi_e&=&-i_m\b F,\\
\dd\psi_m&=&e^{-2\phi}i_m\hd\b F,\label{eq:hd1}
\eq
${}^3\!R$ is the Ricci scalar of the metric $\gamma_{ij}$, and the
 metric $\gamma_{ij}$ has been used to perform the contractions in 
Eq.~(\ref{eq:n4lag}). The Hodge dual in Eq.~(\ref{eq:hd1}) 
is that from the four-dimensional metric (\ref{eq:hd2}).
In order to derive Eq.~(\ref{eq:n4lag}) we have needed to perform a
Legendre transform, which has the effect of changing the sign of the
$|\D\psi_m|^2$ term from what one might have na\"\i{}vely expected.
 We now define new
coordinates
\be
X_+=X^{1/2}e^{\phi}\qquad\mbox{and}\qquad
X_-=X^{1/2}e^{-\phi}.
\ee
Together with the electrostatic potentials 
$\psi_+=\sqrt2\psi_e$ and $\psi_-=\sqrt2\psi_m$. The 
metric on the target space of the harmonic map is given by
\be
G_{AB}\ddd{\phi^A}{\phi^B}=
\frc{\ddd{X_+}{X_+}+\ddd{\psi_+}{\psi_+}}{X_+^2}
+\frc{\ddd{X_-}{X_-}+\ddd{\psi_-}{\psi_-}}{X_-^2}.
\ee
We remark that this precisely takes the form as the sum of two copies of the 
Lagrangian for pure gravity. For the moment we merely note that we can perform independent
 Ehlers' transformations to both $X_+$ and $X_-$ to derive new solutions.

\subsubsection{The Double Ehlers' Transformation}
\label{sect:n4trans}

Performing independent Ehlers' transformations to the system yield 
the following:
\beq
X&\mapsto&\frc X{\left[1+\beta^2\(Xe^{2\phi}+\psi_+^2\)\right]\left[1+\gamma^2\(Xe^{-2\phi}+\psi_-^2\)\right]};\label{eq:stringy1}\\
e^{2\phi}&\mapsto&e^{2\phi}\frc{1+\gamma^2\(Xe^{-2\phi}+\psi_-^2\)}{1+\beta^2\(Xe^{2\phi}+\psi_+^2\)};\\
\psi_+&\mapsto&\frc{\psi_++\beta\(Xe^{2\phi}+\psi_+^2\)}{1+\beta^2\(Xe^{2\phi}+\psi_+^2\)};\\
\psi_-&\mapsto&\frc{\psi_-+\gamma\(Xe^{-2\phi}+\psi_-^2\)}{1+\gamma^2\(Xe^{-2\phi}+\psi_-^2\)}.\label{eq:stringy2}
\eq
In particular if we apply this to Minkowski space we generate 
the Stringy Melvin Universe:
\beq
\b g&=&\(1+\beta^2r^2\sin^2\theta\)\(1+\gamma^2r^2\sin^2\theta\)\(-\ddd tt+\ddd rr+r^2\ddd\theta\theta\)\nonum\\&& \hspace{4cm}{}+\frc{r^2\sin^2\theta\ddd\varphi\varphi}{\(1+\beta^2r^2\sin^2\theta\)\(1+\gamma^2r^2\sin^2\theta\)};\\
e^{2\phi}&=&\frc{1+\gamma^2r^2\sin^2\theta}{1+\beta^2r^2\sin^2\theta};\\
\b A&=&-\sqrt2\gamma r\cos\theta\dd t+\frc{\beta r^2\sin^2\theta\dd\varphi}{\sqrt2\(1+\beta^2r^2\sin^2\theta\)}.
\eq
This solution represents the stringy generalization of Melvin's Universe. Whereas in Melvin's universe
the electric and magnetic fields can be transformed into one another by a simple duality rotation without affecting the metric
(meaning often that we need only consider a purely magnetic or electric universe), the
stringy universe typically involves both electric and magnetic fields. These fields are parallel and provide
a repulsive force to counterbalance the attractive force of the spin zero dilaton and spin two graviton fields.
The Stringy Melvin Universe will be important to us as it will model a strong electromagnetic field
in string theory and one might be interested in the possible mediation of topological defects by such fields.

\section {The Class of Solutions}
\label{sect:cos}

Our starting point is the spherically symmetric solution found by Gibbons \cite{NuclB}:
\be
\b g=-\(1-\frc{2M}r\)\ddd tt+\(1-\frc{2M}r\)^{-1}\!\ddd rr+r\(r-\frc{Q^2}M\)\dd\b\Omega^2.
\label{eq:spherical}
\ee
The electromagnetic field and dilaton are given by
\beq
\b A&=&\frc Qr\dd t
\label{eq:spheric2}
,\\
e^{2\phi}&=&1-\frc {Q^2}{Mr}
\label{eq:spheric3}
\eq
where we write $\phi$ for the dilaton field and $\varphi$ for the angular coordinate.

Let us now  apply the Double Ehlers' Transformation associated with the angular
 Killing vector $\partial/\partial\varphi$. The transformations are given by
 Eqs.~(\ref{eq:stringy1}) to (\ref{eq:stringy2}).

The solution given above, Eqs.~(\ref{eq:spherical}) to (\ref{eq:spheric2}) 
have potentials:
\beq
X&=&r\(r-\frc{Q^2}M\)\sin^2\theta,\\
\psi_+&=&0,\\
\psi_-&=&\sqrt2Q\cos\theta,
\eq
together with (\ref{eq:spheric3}). In consequence it is a simple matter 
to write down the transformed metric and fields:
\beq 
\b g&=&\Lambda\Theta\(-\(1-\frc{2M}r\)\ddd tt+\(1-\frc{2M}r\)^{-1}\ddd rr+r\(r-\frc {Q^2}M\)\ddd\theta\theta\)\nonum\\
&&\hspace{4cm}{}+
\frc r{\Lambda\Theta}\(r-\frc{Q^2}M\)\sin^2\theta\ddd\varphi\varphi,
\eq
where
\be
\Lambda=1+\beta^2\(r-\frc{Q^2}M\)^2\sin^2\theta\qquad\mbox{and}\qquad\Theta=1+\gamma^2r^2\sin^2\theta.
\ee
The new dilaton and potentials are given by
\beq
e^{2\phi}&=&\(1-\frc{Q^2} M\)\frc\Theta\Lambda\, ,\\
\psi_+&=&\frc\beta\Lambda\(r-\frc{Q^2}M\)^2\sin^2\theta,\\
\psi_-&=&\frc1\Theta\left[\sqrt2Q\cos\theta+\gamma\(r^2\sin^2\theta
+2Q^2\cos^2\theta\)\right].
\eq

\section {The Hypotheses}
\label{sect:thyp}

Let us now list the hypotheses we will need to prove our uniqueness theorems:

\begin{itemize}
\item{Axisymmetry: There exists a Killing vector $m$ such that $\lm\b g=0$, $\lm\b F=0$ and $\lm\phi=0$ 
which generates a one-parameter
group of isometries whose orbits are closed spacelike curves.}

\item{Staticity: There exists a hypersurface orthogonal Killing
 vector field $K$ such that $\lk\b g=0$, $\lk\b F=0$ and $\lk\phi$ which generates a one-parameter
group of isometries which acts freely and whose orbits near infinity
are timelike curves.}

\item{Commutivity: $[K,m]=0$.}

\item{Source-free Maxwell equations $\dd\b F=0$ and $\bd\(e^{-2\phi}\b F\)=0$
together with the Einstein equations $R_{ab}=2\D_a\phi\D_b\phi+8\pi e^{-2\phi}T_{ab}^{(F)}$ where
\be
T_{ab}^{(F)}=\frac1{4\pi}\(F_{ac}F_b{}^c-\frac14g_{ab}F_{cd}F^{cd}\).\nonum
\ee}

\item{The domain of outer communication is connected and
simply-connected.}

\item{The solution contains a single black hole.}

\item{The solution is asymptotically the Stringy Melvin Universe.}

\item{Boundary conditions (See section~\ref{sect:n4bcs}).}

\end{itemize}

We remark that the Generalized Papapetrou theorem of as in~\cite{Wells}
 goes through with a few very minor changes to take account of the
 modified Einstein and Maxwell relations. In particular the invariance of
 the dilaton field under the symmetries reads
\be
i_K\dd\phi=0\qquad\mbox{and}\qquad i_m\dd\phi=0.
\ee
Let us define $\b T(\k)=\(T_{ab}-\sh g_{ab}T^c{}_c\)k^a\e b$ be the trace-reverse energy-momentum 1-form with respect to $\k$.
Accordingly the dilaton does not contribute to $\b T(\k)$. In addition the Staticity condition means that the cross term in the metric vanishes, i.e., $W=0$.

The next step is to introduce Weyl coordinates. We show 
that $\rho$ is a harmonic function on the space of orbits. 
Explicitly we have $\rho^2=XV$. Defining
\be
\(h_{AB}\)=\pmatrix{-V&0\cr 0&X}\qquad\mbox{and}\qquad
\(h^{AB}\)=\frc1{\rho^2}\pmatrix{-X&0\cr 0&V},
\ee
 we need to calculate
\be
{}^4\!R_{AB}h^{AB}=-\frc1{2\rho}\D^\alpha\(\rho h^{AB}\D_\alpha h_{AB}\)=-\frc1\rho\D^2\rho.
\ee
Here $A$ and $B$ refer to the $t$ and $\varphi$ coordinates whilst 
the covariant derivatives are with respect to the induced metric on 
the two-dimensional orbit space. Defining
\be
E_\alpha=F_{t\alpha}\qquad\mbox{and}\qquad B_\alpha=F_{\varphi\alpha}
\ee
we have
\beq
{}^4\!R_{tt}&=&e^{-2\phi}\(2\b{E.E}+\sh VF^2\),\\
{}^4\!R_{\varphi\varphi}&=&e^{-2\phi}\(2\b{B.B}-\sh XF^2\).
\eq
where
\be
F^2=2(-X\b{E.E}+V\b{B.B})\rho^{-2}.
\ee
Notice that the invariance of $\phi$ means that $\partial\phi/\partial t=0$
 and $\partial\phi/\partial\varphi=0$, and that therefore $\D_A\phi\D_B\phi$ 
makes no contribution to ${}^4\!R_{AB}$. The result is
\be
-\frc1\rho\D^2\rho=\frc1{\rho^2}\(-{}^4\!R_{tt}X+{}^4\!R_{\varphi\varphi}V\)=0.
\ee
Thus $\rho$ is harmonic and we may go on to introduce its harmonic
 conjugate, $z$ together with $t$ and $\varphi$ that provide a coordinate system for the spacetime.

\subsection{The Divergence Identity}
\label{sect:divid}

We recall at this point our discussion in Sect.~\ref{sect:n4}
 and in particular that the effective two dimensional 
Lagrangian arising from string theory and $N=4$ Supergravity takes the form
\be
{\cal L}=\rho\sqrt{|\gamma|}\left[
\frc{|\D X_+|^2+|\D\psi_+|^2}{X_+^2}+\frc{|\D X_-|^2+|\D\psi_-|^2}{X_-^2}
\right]
\ee
where
\beq
X_+^2=Xe^{2\phi} \qquad\mbox{and}\qquad X_-^2=Xe^{-2\phi}.
\eq
Each term in the above Lagrangian is a copy of the Lagrangian for pure 
gravity and in consequence we may thus use Robinson's identity \cite{Robinson},
\beq
&&\D.\(\rho\D\(\frc{\hat{X_+}^2+\hat{\psi_+}^2}{X_+^{(1)}X_+^{(2)}}+\frc{\hat{X_-}^2+\hat{\psi_-}^2}
{X_-^{(1)}X_-^{(2)}}\)\)\nonum\\&&\hspace{2cm}=
F(X_+^{(1)},X_+^{(2)},\psi_+^{(1)},\psi_+^{(2)})+F(X_-^{(1)},X_-^{(2)},\psi_-^{(1)},\psi_-^{(2)})\ge0,
\eq
where $F(\X1,\X2,\Y1,\Y2)$ is defined by
\beq
&&F(\X1,\X2,\Y1,\Y2)=\frc\rho{\X1\X2}\left[\frc{\hat Y\D\Y1}{\X1}+\frc{\X1\D\X2}
{\X2}-\D\X1\right]^2\nonum\\&&{}\qquad
+\frc\rho{\X1\X2}\left[\frc{\hat Y\D\Y2}{\X2}-\frc{\X2\D\X1}{\X1}+\D\X2\right]^2
\nonum\\&&{}\qquad
+\frc\rho{2\X1\X2}\left[\(\frc{\D\Y2}{\X2}-\frc{\D\Y1}{\X1}\)(\X1+\X2)-
\(\frc{\D\X2}{\X2}+\frc{\D\X1}{\X1}\)\hat Y\right]^2
\nonum\\&&{}\qquad
+\frc\rho{2\X1\X2}\left[\(\frc{\D\Y2}{\X2}+\frc{\D\Y1}{\X1}\)\hat X-
\(\frc{\D\X2}{\X2}+\frc{\D\X1}{\X1}\)\hat Y\right]^2.
\eq

We have defined ${\df A}=A_2-A_1$ etc.
It is now evident that we may use this divergence identity to provide us with 
the key tool in establishing a black hole uniqueness theorem. To complete
the proof
we will want to change coordinates, and impose suitable boundary
 conditions to make the relevant boundary integral vanish. 
We make the change of coordinates:
\beq
\rho&=&r\sin\theta,\\
z&=&r\cos\theta.
\eq
The value of $r$ runs from $M$ to infinity (we adjust the additive 
constant to $z$ to make the horizon run from $-M\le z\le M$). The overall 
scaling of $\rho$ and $z$ is made such that asymptotically $r$ becomes the 
radial coordinate of the Stringy Melvin Universe, i.e.,
\be
\b g\sim A\rho^4(-\ddd tt+\ddd\rho\rho+\ddd zz)
+\frc1{A\rho^2}\ddd\varphi\varphi.
\ee
with $\varphi$ taking values in $[0,2\pi)$. It is worth remarking that
we cannot rescale the coordinates
and parameters and retain this form whilst leaving the range of $\varphi$ 
unchanged, except for the trivial instance of multiplying the coordinates by $-1$. 

The two dimensional domain we work on is the semi-infinite rectangle, $r>M$ and
 $-\pi/2\le\theta\le\pi/2$, and the boundary integral we require to vanish is
 now given by
\be
\int r\cos\theta
 \(r d\theta\ppp r-\frc{dr}r\ppp\theta\)
\(\frc{\hat{X_+}^2+\hat\psi_+^2}{X_+^{(1)}X_+^{(2)}} +\frc{\hat{X_-}^2+\hat\psi_-^2}{X_-^{(1)}X_-^{(2)}}\)=0.
\ee

\subsection{Boundary Conditions}
\label{sect:n4bcs}

We now need to impose suitable boundary conditions to make the 
boundary integral vanish. The following prove to be sufficient.
At infinity we require
\beq
X_+&=&\frc1{\beta^2\sin\theta}\,\frc1{r}+O\(\frc1{r^2}\);\\
\frc1{X_+}\pp {X_+}r&=&-\frc1r+O\(\frc1{r^2}\);\\
\psi_+&=&\frc1{\beta}-\frc1{\beta^3\sin^2\theta}\,\frc1{r^2}+O\(\frc1{r^3}\);\\
\pp{\psi_+} r&=&\frc2{\beta^3\sin^2\theta}\,\frc1{r^3}+O\(\frc1{r^4}\);\\
X_-&=&\frc1{\gamma^2\sin\theta}\,\frc1{r}+O\(\frc1{r^2}\);\\
\frc1{X_-}\pp {X_-}r&=&-\frc1r+O\(\frc1{r^2}\);\\
\psi_-&=&\frc1{\gamma}-\frc{1+\sqrt2\gamma Q\cos\theta}{\gamma^3\sin^2\theta}\,\frc1{r^2}+O\(\frc1{r^3}\);\\
\pp{\psi_-} r&=&\frc{2\(1+\sqrt2\gamma Q\cos\theta\)}
{\gamma^3\sin^2\theta}\, \frc1{r^3}+O\(\frc1{r^4}\).
\eq
On the axes we require (setting $\mu=\sin\theta$)
\beq
\frc1{X_+}\pp{X_+}\mu&=&\frc{-\mu}{1-\mu^2}+O(1);\\
\pp{X_+}r&=&O\(\(1-\mu^2\)^{1/2}\);\\
\psi_+&=&O(1-\mu^2);\\
\pp{\psi_+}\mu&=&O(1);\\
\pp{\psi_+} r&=&O(1-\mu^2);\\
\frc1{X_-}\pp{X_-}\mu&=&\frc{-\mu}{1-\mu^2}+O(1);\\
\pp{X_-}r&=&O\(\(1-\mu^2\)^{1/2}\);\\
\psi_-&=&\frc{(\sqrt2\mu+2\gamma Q)Q}{1+2\gamma^2 Q^2}
+O(1-\mu^2);\\
\pp{\psi_-}\mu&=&O(1);\\
\pp{\psi_-} r&=&O(1-\mu^2);
\eq
where the boundaries correspond to $\mu=\pm1$. On the horizon we require regularity
 of $X_+$, $X_-$, $\psi_+$ and $\psi_-$. These conditions are sufficient to make 
the boundary integral vanish and hence establish our uniqueness result.

\section{Uniqueness Theorems for the Stringy $C$-metric and 
Stringy-Ernst Solution}
\label{sect:stringyunique}

In \cite{Wells} we proved the uniqueness of both the $C$-metric and  the
 Ernst solution. In this section we exploit the techniques developed there
 together with the string uniqueness formalism we have just been using to 
show that given any Stringy $C$-metric or Stringy Ernst solution then the
 boundary conditions uniquely specify the solution. Our philosophy here is
 slightly less ambitious than for Einstein-Maxwell theory; in the latter case
 we took the position that any candidate solution that resembled the Ernst 
solution at infinity was indeed an Ernst solution provided one of the quantities
 determined on the boundary was greater than a critical value. Here we
 assume we have an Ernst solution that does satisfy the boundary conditions
 and prove that no other solution can have the same boundary conditions.

Our starting point is the Dilaton $C$-metric found by Dowker {\em et al.\/} 
\cite{Dowker},
\beq
\b g&=&\frc1{A^2(x-y)^2}\left[F(x)G(y)\ddd tt+\frc{F(y)\ddd xx}{G(y)}-\frc{F(x)\ddd yy}{G(y)}\right.\nonum\\
&&{}\hspace{5cm}+F(y)G(x)\ddd\varphi\varphi\Bigg] ,
\eq
where
\beq
e^{-2\phi}&=&\frc{F(y)}{F(x)}\, ,\\
\b A&=&\sqrt{\frc{r_+r_-}2}(x-x_2)\dd\varphi,\\
F(\xi)&=&1+r_-A\xi\, ,\\
G(\xi)&=&1-\xi^2-r_+A\xi^3.
\eq
We have labelled the roots of $G(x)$ as $x_3<x_2<x_1$ with $x_1>0$.
 The quantity $x_4$ corresponds to setting $F(x)=0$, for which we assume
 $x_4<x_3$ so as to represent an inner horizon for the black hole.

It is advantageous to represent this solution in terms of the Jacobi 
elliptic functions. We transform to new coordinates using
\be
\frc\chi M=\int_{x_2}^x\frc{d\xi}{\sqrt{F(\xi)G(\xi)}}\qquad\mbox{and}\qquad
\frc\eta M=\int_y^{x_2}\frc{d\xi}{\sqrt{-F(\xi)G(\xi)}}.
\ee
with $M=\sqrt{e_1-e_3}$ where $e_i=\wp(\omega_i)$ and $\omega_i$ being a
 half period as we had in Appendix~A of \cite{Wells}. The appropriate 
invariants of the $\wp$-function are given by
\beq
g_2&=&\frc{1+3A^2r_-^2-9A^2r_+r_-}{12}, \\
g_3&=&\frc{2-27A^2r_+^2-18A^2r_-^2+27A^2r_+r_-+27A^4r_+r_-^3}{432}.
\eq
Writing the metric as
\be
\b g=-V\ddd tt+X\ddd\phi\phi+\Sigma\(\ddd\chi\chi+\ddd\eta\eta\),
\ee
we find:
\beq
X&=&\frc{4L^2\(1-D\sn^2\eta\)\(1-E\sn^2\eta\)\sn^2\chi\cn^2\chi\dn^2\chi}
{\(\cn^2\chi+D\sn^2\chi\)\(\cn^2\eta+E\sn^2\eta\)\(\sn^2\chi+\sn^2\eta\cn^2\chi\)^2}\, ;\\
V&=&\frc{4L^2\(\cn^2\chi+D\sn^2\chi\)\(\cn^2\chi+E\sn^2\chi\)\sn^2\eta\cn^2\eta\dn^2\eta}
{\(1-D\sn^2\eta\)\(1-E\sn^2\eta\)\(\sn^2\chi+\sn^2\eta\cn^2\chi\)^2}\, ;\\
\Sigma&=&\frc{16H^2L^2\(\cn^2\chi+D\sn^2\chi\)\(\cn^2\chi+E\sn^2\chi\)\(1-D\sn^2\eta\)^2\(1-E\sn^2\eta\)}
{\kappa^2\(\sn^2\chi+\sn^2\eta\cn^2\chi\)^2}\nonum\, .\\
\ 
\eq
We have written $M=AL$ together with
\be
\kappa=\left.\frc{d(F(\xi)G(\xi))}{d\xi}\right|_{\xi=x_2}\ ,\qquad 
D=\frc{1+k'^2}3-\frc1{24M^2}\,\left.\frc{d^2(F(\xi)G(\xi))}{d\xi^2}
\right|_{\xi=x_2}
\ee
and
\be
E=D+\frc{r_-A\kappa}{4M^2H}\ ,\qquad H=1+Ar_-x_2.
\ee
The quantity $\rho$ is given by
\be
\rho=\frc{4L^2\sn\chi\cn\chi\dn\chi\sn\eta\cn\eta\dn\eta}
{\(\sn^2\chi+\sn^2\eta\cn^2\chi\)^2}.
\ee
Thus we have $z-i\rho=2L^2\wp(\chi+i\eta)$. The dilaton and vector potential are given by the expressions
\beq
e^{-2\phi}&=&\frc{\(\cn^2\chi+D\sn^2\chi\)\(1-E\sn^2\eta\)}
               {\(\cn^2\chi+E\sn^2\chi\)\(1-D\sn^2\chi\)}\, ;\\
\b A&=&\frc{QD\sn^2\chi\dd\varphi}{4\(\cn^2\chi+D\sn^2\chi\)}\, ;
\qquad Q=\frc{\kappa\sqrt{r_+r_-}}{\sqrt2A^3L^2D}.
\eq
Performing the transformations Eqs.~(\ref{eq:stringy1}) to (\ref{eq:stringy2}) we arrive at
the metric of interest. The new metric and fields we have derived are:
\beq
X&=&\frc{4L^2\(1-D\sn^2\eta\)\(1-E\sn^2\eta\)\sn^2\chi\cn^2\chi\dn^2\chi}
{\Lambda\Theta\(\cn^2\chi+D\sn^2\chi\)\(\cn^2\eta+E\sn^2\eta\)
\(\sn^2\chi+\sn^2\eta\cn^2\chi\)^2}\, ;\\
V&=&\frc{4L^2\Lambda\Theta\(\cn^2\chi+D\sn^2\chi\)\(\cn^2\chi+E\sn^2\chi\)\sn^2\eta\cn^2\eta\dn^2\eta}
{\(1-D\sn^2\eta\)\(1-E\sn^2\eta\)\(\sn^2\chi+\sn^2\eta\cn^2\chi\)^2};\\
\Sigma&=&\frc{16H^2L^2\Lambda\Theta\(\cn^2\chi+D\sn^2\chi\)\(\cn^2\chi+E\sn^2\chi\)\(1-D\sn^2\eta\)^2
\(1-E\sn^2\eta\)}{\kappa^2\(\sn^2\chi+\sn^2\eta\cn^2\chi\)^2};\nonum\\
\ 
\eq
with
\beq
\Lambda&=&1+\beta^2\left\{\frc{4L^2\sn^2\chi\cn^2\chi\dn^2\chi\(1-D\sn^2\eta\)^2}
{\(\cn^2\chi+D\sn^2\chi\)^2\(\sn^2\chi+\sn^2\eta\cn^2\chi\)^2}
+\frc{Q^2D^2\sn^4\chi}{16\(\cn^2\chi+D\sn^2\chi\)}\right\};\nonum\\ \ \\
\Theta&=&1+\frc{4\gamma^2L^2\sn^2\chi\cn^2\chi\dn^2\chi\(1-E\sn^2\eta\)^2}
{\(\cn^2\chi+E\sn^2\chi\)^2\(\sn^2\chi+\sn^2\eta\cn^2\chi\)^2}\ .
\eq
The dilaton is given by
\be
e^{-2\phi}=\frc{\Lambda\(\cn^2\chi+D\sn^2\chi\)\(1-E\sn^2\eta\)}
               {\Theta\(\cn^2\chi+E\sn^2\chi\)\(1-D\sn^2\chi\)}\ .\\
\ee
We record the values of the quantities $X_\pm$ and the potentials $\psi_\pm$:
\beq
X_+&=&\frc{2L\sn\chi\cn\chi\dn\chi\(1-D\sn^2\eta\)}
{\Lambda\(\cn^2\chi+D\sn^2\chi\)\(\sn^2\chi+\sn^2\eta\cn^2\chi\)};\\
X_-&=&\frc{2L\sn\chi\cn\chi\dn\chi\(1-E\sn^2\eta\)}
{\Theta\(\cn^2\chi+E\sn^2\chi\)\(\sn^2\chi+\sn^2\eta\cn^2\chi\)};\\
\psi_+&=&\frc1\Lambda\left\{\frc{QD\sn^2\chi}{4\(\cn^2\chi+D\sn^2\chi\)}
\right.\hspace{6cm}\nonum\\&&{}\left.\hspace{-1.5cm}
+\beta\left[\frc{4L^2\sn^2\chi\cn^2\chi\dn^2\chi\(1-D\sn^2\eta\)^2}
{\(\cn^2\chi+D\sn^2\chi\)^2\(\sn^2\chi+\sn^2\eta\cn^2\chi\)^2}
+\frc{Q^2D^2\sn^4\chi}{16\(\cn^2\chi+D\sn^2\chi\)}\right]\right\};\qquad\\
\psi_-&=&
\frc{4\gamma L^2\sn^2\chi\cn^2\chi\dn^2\chi\(1-E\sn^2\eta\)^2}
{\Theta\(\cn^2\chi+E\sn^2\chi\)^2\(\sn^2\chi+\sn^2\eta\cn^2\chi\)^2}.
\eq

 We will be interested in the behaviour of the fields as one takes the limits
 $\chi\rightarrow0$, $u\rightarrow0$ with $u=K-\chi$ and $R\rightarrow\infty$.
 The appropriate boundary conditions we need to make the boundary integral
 vanish are presented in the next section.

\subsection{Boundary Conditions for the Stringy Ernst Solution and $C$-Metric}
\label{sect:stringybcs}

In order to complete the proof of the uniqueness for the Stringy Ernst solution and
Stringy $C$-metric it only remains to write down a set of boundary conditions
that will make the boundary integral vanish. It is fairly simple to verify that
the conditions given in the following two subsections are sufficient for this
purpose.

\subsubsection{Boundary Conditions for the Stringy 
Ernst Solution Uniqueness Theorem}

To start with we will require all the fields to be regular 
(and in addition for $X_+$ and $X_-$ to not vanish)
as one approaches the acceleration and event horizons. 
Near the axis $\chi=0$ we demand
\beq
\frc1{X_+}\pp{X_+}\chi&=&\frc1\chi+O(1);\\
\pp{X_+}\eta&=&O\(\chi\);\\
\psi_+&=&O\(\chi^{2}\);\\
\pp{\psi_+}\chi&=&O\(\chi\);\\
\pp{\psi_+}\eta&=&O\(\chi\);\\
\frc1{X_-}\pp{X_-}\chi&=&\frc1\chi+O(1);\\
\pp{X_-}\eta&=&O\(\chi\);\\
\psi_-&=&O\(\chi^2\);\\
\pp{\psi_-}\chi&=&O\(\chi\);\\
\pp{\psi_-}\eta&=&O\(\chi\).
\eq
For the other axis we will require
\beq
\frc1{X_+}\pp{X_+}u&=&\frc1u+O(1);\\
\pp{X_+}\eta&=&O\(u\);\\
\psi_+&=&\frc{Q\(4+\beta Q\)}{16+\beta^2 Q^2}+O\(u^2\);\\
\pp{\psi_+}u&=&O\(u^2\);\\
\pp{\psi_+}\eta&=&O\(u\);\\
\frc1{X_-}\pp{X_-}u&=&\frc1u+O(1);\\
\pp{X_-}\eta&=&O\(u\);\\
\psi_-&=&O\(u^2\);\\
\pp{\psi_-}u&=&O\(u\);\\
\pp{\psi_-}\eta&=&O\(u\).
\eq
Whilst as $R\rightarrow\infty$ with $\chi=R^{-1/2}\sin\theta$ and $\eta=R^{-1/2}\cos\theta$ we will demand
\beq
X_+&=&\frc1{2\beta^2L\sin\theta}\,\frc1{R^{1/2}}+O\(\frc1{R^{3/2}}\);\\
\frc1{X_+}\pp{X_+}R&=&-\frc1{2R}+O\(\frc1{R^2}\);\\
\pp{X_+}\theta&=&O\(\frc1{R^{1/2}}\);\\
\psi_+&=&\frc1{\beta}-\frc1{4\beta^3L^2\sin^2\theta}\,\frc1R
+O\(\frc1{R^2}\);\\
\pp{\psi_+}R&=&\frc1{4\beta^3L^2\sin^2\theta}\,\frc1{R^2}
+O\(\frc1{R^3}\);\\
\pp{\psi_+}\theta&=&O\(\frc1R\);\\
X_-&=&\frc1{2\gamma^2L\sin\theta}\,\frc1R+O\(\frc1{R^2}\);\\
\frc1{X_-}\pp{X_-}R&=&-\frc1{2R}+O\(\frc1{R^2}\);\\
\pp{X_-}\theta&=&O\(\frc1{R^{1/2}}\);\\
\psi_-&=&\frc1{\gamma}-\frc1{4\gamma^3L^2\sin^2\theta}\,\frc1R
+O\(\frc1{R^2}\);\\
\pp{\psi_-}R&=&\frc1{4\sqrt2\gamma^3L^2\sin^2\theta}\,\frc1{R^2}
+O\(\frc1{R^3}\);\\
\pp{\psi_-}\theta&=&O\(\frc1{R^{1/2}}\).
\eq
These boundary conditions are sufficient to establish the uniqueness
 of the Stringy Ernst solutions. For good measure we also present the boundary 
conditions for the Stringy $C$-metric problem.

\subsubsection{Boundary Conditions for the Stringy $C$-Metric Uniqueness Theorem}

The appropriate conditions are as follows. Near $\chi=0$ we will insist
\beq
\frc1{X_+}\pp{X_+}\chi&=&\frc1\chi+O(1);\\
\pp{X_+}\eta&=&O\(\chi\);\\
\psi_+&=&O\(\chi^2\);\\
\pp{\psi_+}\chi&=&O\(\chi\);\\
\pp{\psi_+}\eta&=&O\(\chi\);\\
\frc1{X_-}\pp{X_-}\chi&=&\frc1\chi+O(1);\\
\pp{X_-}\eta&=&O\(\chi\);\\
\psi_-&=&O\(\chi^2\);\\
\pp{\psi_-}\chi&=&O\(\chi\);\\
\pp{\psi_-}\eta&=&O\(\chi\).
\eq
For the other axis we will require
\beq
\frc1{X_+}\pp{X_+}u&=&\frc1u+O(1);\\
\pp{X_+}\eta&=&O\(u\);\\
\psi_+&=&\frc Q{2\sqrt2}+O\(u^2\);\\
\pp{\psi_+}u&=&O\(u\);\\
\pp{\psi_+}\eta&=&O\(u\);\\
\frc1{X_-}\pp{X_-}u&=&\frc1u+O(1);\\
\pp{X_-}\eta&=&O\(u\);\\
\psi_-&=&O\(u^2\);\\
\pp{\psi_-}u&=&O\(u\);\\
\pp{\psi_-}\eta&=&O\(u\).
\eq
Whilst as $R\rightarrow\infty$ with $\chi=R^{-1/2}\sin\theta$ 
and $\eta=R^{-1/2}\cos\theta$ we will demand
\beq
X_+&=&2L\sin\theta\,R^{1/2}+O\(1\);\\
\frc1{X_+}\pp{X_+}R&=&\frc1{2R}+O\(\frc1{R^2}\);\\
\pp{X_+}\theta&=&O\(R^{1/2}\);\\
\psi_+&=&\frc{QD\sin^2\theta}{2\sqrt2}\,\frc1R+O\(\frc1{R^2}\);\\
\pp{\psi_+}R&=&-\frc{QD\sin^2\theta}{2\sqrt2}\,\frc1{R^2}+O\(\frc1{R^3}\);\\
\pp{\psi_+}\theta&=&O\(\frc1R\);\\
X_-&=&2L\sin\theta\,R^{1/2}+O\(1\);\\
\frc1{X_-}\pp{X_-}R&=&\frc1{2R}+O\(\frc1{R^2}\);\\
\pp{X_-}\theta&=&O\(R^{1/2}\);\\
\psi_-&=&O\(\frc1{R^2}\);\\
\pp{\psi_-}R&=&O\(\frc1{R^3}\);\\
\pp{\psi_-}\theta&=&O\(\frc1R\).
\eq

\subsection{Conclusion}
\label{sect:stringyconc}

We have been able to prove the uniqueness of two classes of asymptotically 
Melvin black holes. We would hope that the formalism developed in this chapter 
to prove the uniqueness of our class of black holes could be used to prove the
 uniqueness of other classes of static solutions in these theories. 
We would also like to have a formalism that incorporates the possibility of rotation 
and includes the axionic field, however it seems likely that such an
extension would not be straightforward. The crux of the uniqueness proof is
to establish the positivity of a suitable divergence. It turned out
that for the static truncation of string theory that we considered the 
Lagrangian split into two separate copies of that for pure gravity. Consequently
we could simply add together two copies of the relevant divergence identity
(Robinson's identity) to furnish us with an expression that we could use in
our black hole uniqueness investigations. If we include rotation or an axionic
field the Lagrangian will not decompose so easily, and we would need to
deal with it as a whole. This is problematical as the target space of the
harmonic map possesses (at least) two timelike directions. Unfortunately this
prohibits a simple application of the Mazur construction or a suitable
analogue of the construction presented in \cite{Wells}. It seems
that Bunting's approach may be the best way forward under these circumstances
relying, as it does, more heavily on the negative curvature of the target space metric than
on its particular form as an $SU(1,2)/S\(U(1)\times U(2)\)$ symmetric space
harmonic mapping system.

\end{document}